# An Exploratory Study of Critical Factors Affecting the Efficiency of Sorting Techniques (Shell, Heap and Treap)

**Olusegun Folorunso, Olufunke R. Vincent, Oluwatimilehin Salako**
Department of Computer Science,
University of Agriculture, Abeokuta, Nigeria
folorunsolusegun@yahoo.com

**ABSTRACT.** The efficiency of sorting techniques has a significant impact on the overall efficiency of a program. The efficiency of Shell, Heap and Treap sorting techniques in terms of both running time and memory usage was studied, experiments conducted and results subjected to factor analysis by SPSS. The study revealed the main factor affecting these sorting techniques was time taken to sort.

## Introduction

Sorting is an operation that segregates items into groups according to a specified criterion. It is any process of arranging items in some sequence and/or in different sets, and accordingly. Sorting indicates the distribution of grain size of sediments, either in unconsolidated deposits or in sedimentary rocks. Poorly sorted indicates that the sediment sizes are mixed (large variance); whereas well sorted indicates that the sediment sizes are similar (low variance). It also means ordering- arranging items of the same kind, class, nature, etc. This ordering makes this possible or easy to search a specific data element from sorted elements [CKY06]. In computer science sorting is very much important. As efficiency is a major concern of computing, thus data is sorted in order to gain the efficiency in retrieving or searching tasks.

Sorting allows information to be put into a meaningful order. In any database management system, sorting plays a big role [R+09]. All the data in a database, is stored in a sorted and compact form by a database management system. However, sorting techniques takes a very long time to sort large sets of data. Sorting algorithms put elements of a list in a certain order mostly numerical or lexographical. Efficient sorting is important to optimizing the use of other algorithms such as search and merge algorithms that requires sorted list to work correctly [Mil09]. Different ideas which lead to sorting algorithms are insertion, exchange, merging selection, merging.

This paper aim at studying the efficiency with which sorting is carried using shell, heap and treap sorting techniques. The efficiency of the three techniques will be analysed in terms of both running time and memory usage. The sorting time taken, memory consumed and total memory used will be used in this study as decision variables to evaluate their efficiencies.

## 1. Background to Sorting Techniques

The oldest technique to sorting is the shell sorting which was proposed by Shell [She59] and still in many cases holds its own against other competitors due to its simplicity and the ability to use partially ordered sequences. It is considered to be an elegant extension of insertion sort that gains speed by allowing exchanges of elements that are far apart. It sorts slices with a particular step h. Such a file is said to be h-sorted. If we first h-sort a file using a large value h, elements move distances and the efficiency of *h*-sorting for smaller values of h improves. When the value of *h* equals 1, we implement a regular insertion sort and thus get a sorted file.

The heap sort works as its name suggests. It begins by building a heap out of the data set, removes the largest item and places it at the end of the sorted array. After removing the largest item, it reconstructs the heap and removes the largest remaining item and places it in the next open position from the end of the sorted array. This is repeated until there are no items left in the heap and the sorted array is full. Elementary implementations require two arrays – one to hold the heap and the other to hold the sorted elements [Wil64].

In treap sort, each node stores a key and a priority. The treap is a binary search tree with respect to the keys and a heap with respect to the priorities. The treap corresponding to a given set of key-priority pairs is unique. Insertions into treaps are performed like in standard binary search trees, followed by a sequence of rotations that restore the heap property. To delete an element, we rotate it down to the bottom of the tree and then remove it. The interesting property of treaps is that if the priorities are chosen uniformly at random from the interval, then all the treap operations take expected time *0 (log n)* .

Shell, Heap and Treap sorting algorithms have their strengths and weaknesses as to running time and system resources consumed. Three factors, the time taken to execute the sort, the memory consumed and the total memory used were considered in evaluating the three sorting techniques. This paper aims at determining the most critical of the three factors. Experimental results for the decision variables were generated from an algorithm implemented in java in which the amount of numbers sorted were varied for the three different sorting techniques. Factor analysis by principal components of the obtained experimental data was carried out using Statistical Package for Social Scientists (SPSS) for the purpose of estimating the contribution of each factor to the success of the sorting algorithms and one factor was extracted. Further statistical analysis was carried out to generate eigenvalue of the extracted factor. The eigenvalue forms the basis for estimating the contribution of the extracted factor. Moreover, a system of linear equations which can be used to estimate the assessment of each assessor of the sorting techniques is proposed.

## 2. Materials and Methods

The decision variables of the impact of time taken, memory consumed and total memory used relate to one another. The general form of the mathematical model for evaluating the decision variables is presented as:

$$Y_i = \sum_{k=1}^{n} a_{i,k} - X_k \quad ...i = 1,2,3,...,m \qquad \text{eqn (1)}$$

Where $Y_i$ represents the $i^{th}$ assessor's observation of decision variable $X_k$; $a_{i,k}$ represents the assessment of $k^{th}$ decision variable by $i^{th}$ Assessor.
This mathematical model can be expressed as:

$$\begin{bmatrix} Y_1 \\ Y_2 \\ \vdots \\ Y_m \end{bmatrix} = \begin{bmatrix} a_{1,1} & X_1 + & \cdots & + a_{1,3} & X_3 \\ \vdots & & \ddots & & \vdots \\ a_{m,1} & X_1 + & \cdots & + a_{m,3} & X_3 \end{bmatrix} \quad \text{(eqn 2)}$$

The factor analysis by principal components is adopted in the evaluation of the decision variable of the impact of time. The primary goal is to obtain the contribution of each of the factors to the efficiency of the sorting techniques. The following statistics were generated and used for the above stated objective: Descriptive statistics, Correlation matrix, Bartlett's test and Kaiser-Mayer Olkin (KMO), Communalities, Initial factor loadings, Rotated factor loadings, Factor score coefficient matrix, Eigen values.

The descriptive statistics presents the mean and standard deviation of the raw score of each performance indices given by the sample Assessors. The correlation matrix presents the degree of pair wise relationships of the performance indices. The Bartlett's test of sphericity is used to test the adequacy of the sample population. Another measure of the adequacy of sample population is Kaiser-Mayer Olkin (KMO).

In factor analysis, there is a set of factors which is generally referred to as 'common factors' each of which loads on some performance indices and another set of factors which are extraneous to each of the performance indices. The proportion of a variance of a performance indices explained by the common factor is called the 'communality' of the performance indices. The communality of the performance index ranges between 0 and 1, where 0 indicates that the common factors explains none of the variance and 1 indicates that all the variance is explained by the common factors.

According to [AAU09], the component matrix presents the initial factor loadings. The factor loadings associated with a specific index is simply the correlation between the factor and the standard score of the index. The degree of generalization found between each index and each factor is referred to as 'factor loading'. The farther away a factor loading is from zero in the positive direction, the more one can conclude the contribution of an index to a factor. The component matrix can be rotated by varimax, promax, equamax or quartimax for the purpose of establishing a high correlation between indices and factors. The factor score coefficient matrix can be used to evaluate the assessment of each Assessor is generated. In compliance with earlier researchers [AAU09], the eigenvalues and percentage variance of the factors considered are generated, as well, for the purpose of evaluating the contributions of each factor to the efficiency of the sorting techniques.

### 2.1. Data Collection, Analysis and Interpretation of Results

Algorithms implemented in java were used to generate experimental data for the three various sorting techniques. The number of values to sort was varied for the three sorting techniques which produced different results for the 'time taken', 'memory consumed' and 'total memory used'. For the shell algorithm the input is an array *a* of length *n* while the output is a sorted array. The input for the Heap Algorithm is an almost complete binary tree with root *r* and vertex labeling *a* while the output is an ordered/sorted Array. It uses the procedure of buildheap() and downheap(). The treap algorithm finds insertion point and the *k* and continues until sorting is achieved.

### 2.2. Data Generated

The descriptive statistics of the data collected exhibits the mean and standard deviation of the rating of the impact of time and memory on the efficiency of the sorting techniques by the experimental results generated. For example, the mean and standard deviation of the rating on memory consumed in bits for shell sort are 1080681.33 and 170074.128 respectively. For Treap, the mean and standard deviation of the rating on memory consumed in bits is 1265118.00 and 43517.031 respectively and 1349945.33 and 111522.008 for that of Heap. Thereafter, the final data were subjected to factor analysis by principal components using SPSS package.

The extraction method was by principal component analysis and the rotation method promax with Kaiser Normalization. According to the computed analysis, Heap sort for instance shows that the correlation of 0.172 exists between 'memory consumed' and 'total memory used'. The correlation of 0.869 exists between 'memory consumed' and 'time taken' to sort. The implication is that 'memory consumed' is not likely to share the same factor with 'total memory used'. On the other hand, 'memory consumed' is very likely to share the same factor with 'time taken' to sort. The Bartlett's test for Heap sort for instance produces a $X^2$ of 140.036, degree of freedom of 3 and a significance level of 0.000, which indicates the adequacy of the sample data. The results obtained from the Bartlett's test and KMO test are good indicators of the suitability of factor analysis as well.

The communalities of the performance indices generated for the sorting techniques with principal component analysis as the extraction method are presented in tables 1 through 5, with initial values for all three factors (time taken (nano second), memory consumed (bits), total memory used (kb)) considered taken as 1.000 for heap, treap and shell methods.

**Table 1: Component Score Coefficient Matrix for Heap**

|  | Component | | |
|---|---|---|---|
|  | 1 | 2 | 3 |
| Time taken(nano second) | -.699 | 1.671 | -.815 |
| Memory Consumed(bits) | .822 | -.344 | 438.933 |
| Total memory used(KB) | .849 | -.355 | -437.960 |

**Table 2: Component Score Coefficient Matrix for Treap**

|  | Component | | |
|---|---|---|---|
|  | 1 | 2 | 3 |
| Time taken(nano second) | -.602 | 1.575 | -.657 |
| Memory Consumed(bits) | .813 | -.311 | -184.432 |
| Total memory used(KB) | .763 | -.292 | 185.244 |

**Table 3: Component Score Coefficient Matrix for Shell**

|  | Component | | |
|---|---|---|---|
|  | 1 | 2 | 3 |
| Time taken(nano second) | -.141 | 1.122 | -.370 |
| Memory Consumed(bits) | .555 | -.070 | 450.224 |
| Total memory used(KB) | .567 | -.071 | -449.419 |

The generated component score coefficient matrices are used to estimate the assessment of each assessor of the impact of time and memory on the efficiency of the sorting techniques.

This can be achieved by formulating a linear equation of the form:

$$C_{i,j} = \sum_{k=1}^{3} b_{k,j} S_{i,k} \quad i = 1, 2, \ldots n; \quad j = 1 \qquad \text{Eqn (3)}$$

Where $C_{i,j}$ represents the contribution of *ith* assessor to $j^{th}$ factor; $b_{k,j}$ represents the component score coefficient of k*th* decision variable for $j^{th}$ factor; $S_{i,k}$ represents the standard score of *ith* assessor for $k^{th}$ decision variable and *n* represents the number of sampled assessors.
$S_{i,k}$ is estimated by:

$$S_{i,k} = A + \frac{(x_i + y_i)}{d_i} \qquad \text{Eqn (4)}$$

Where *A* represents the allowable minimum raw score for decision variable; in this instance, it is *1*; $x_i$ represents the raw score of *ith* decision variable; $y_i$ represents the mean of the raw scores of *ith* decision variable; $d_i$ represents the standard deviation of the raw scores of $i^{th}$ decision variable. For each sampled Assessor, the system of linear equations for the single extracted factor can be represented as follows:

$$b_{1,1}S_{i,1} + b_{2,1}S_{i,2} + ... + b_{4,1}S_{i,4} = C_{i,1} \qquad \text{Eqn (5)}$$

In an attempt to evaluate the percentage contribution of each factor to the efficiency of the sorting techniques, the eigen value of each factor is generated. The eigen value of $j^{th}$ factor denoted by '$E_j$' is calculated by:

$$E_j = \sum_{k=1}^{3} X^2_{i,j} \qquad i = 1, 2, 3; \quad j = 1 \qquad \text{Eqn (6)}$$

Where $X_{i,j}$ represents the loading of $j^{th}$ factor on ith decision variable.
The eigenvalue is used to indicate how well each of the factors fits the experimental data. The percentage

$$P = 100 \left(\frac{E_j}{n}\right) \qquad \text{Eqn (7)}$$

Where *n* represents the number of decision variables considered in our study. Tables 4 to 6 present the eigenvalues, the percentage contribution and cumulative percentage contribution of the three considered factors for each of the three sorting techniques according to [AAU09].

**Table 4: Eigen value generated for Heap**

| Component | Initial Eigenvalues | | | Rotation Sums of Squared Loadings |
|---|---|---|---|---|
| | Total | % of Variance | Cumulative % | Total |
| 1 | 2.969 | 98.966 | 98.966 | 2.826 |
| 2 | .031 | 1.034 | 100.000 | 2.720 |
| 3 | 2.519E-6 | 8.396E-5 | 100.000 | .094 |

**Table 5: Eigen value generated for Treap**

| Component | Initial Eigenvalues | | | Rotation Sums of Squared Loadings |
|---|---|---|---|---|
| | Total | % of Variance | Cumulative % | Total |
| 1 | 2.962 | 98.748 | 98.748 | 2.822 |
| 2 | .038 | 1.251 | 100.000 | 2.704 |
| 3 | 1.419E-5 | .000 | 100.000 | .090 |

**Table 6: Eigen value generated for Shell**

| Component | Initial Eigenvalues | | | Rotation Sums of Squared Loadings |
|---|---|---|---|---|
| | Total | % of Variance | Cumulative % | Total |
| 1 | 2.840 | 94.677 | 94.677 | 2.705 |
| 2 | .160 | 5.323 | 100.000 | 2.425 |
| 3 | 1.854E-6 | 6.178E-5 | 100.000 | .699 |

The three factors contribute a total of 100% to the efficiency of the three sorting techniques. From the results, 'time taken' contributed 98.966% and 'memory consumed' contributed 1.034% impact on the efficiency of Heap sorting technique. This can be visualized in Figure 1.

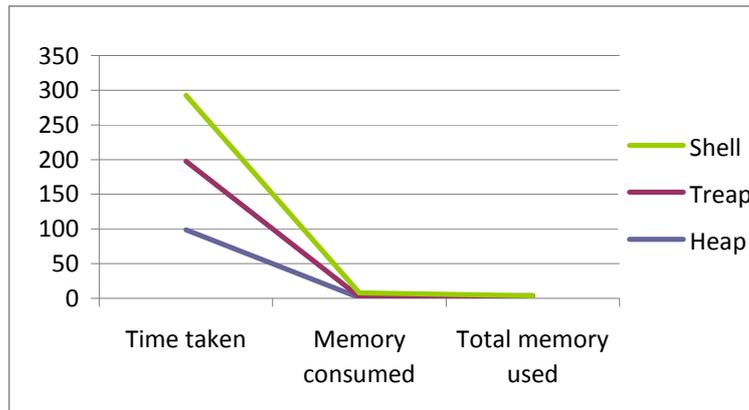

**Figure 1: Comparison of sorting techniques based on factors considered**

## Concluding Remarks and Future Work

The efficiency with which sorting is carried out often has a significant impact on the overall efficiency of a program. The efficiency of Shell, Heap and Treap sorting techniques in terms of both running time and memory usage was studied, experiments conducted and results subjected to factor analysis by SPSS. The sort time taken, memory consumed and total memory used was used as decision variables to evaluate their efficiencies. Experimental results for the decision variables were generated from a software tool in which the amount of numbers sorted were varied for the three different sorting techniques. The results were subjected to factor analysis using Statistical Package for Social Scientists (SPSS) to test the level at which each of the factors affect the sorting techniques. Eigen values were used to indicate how well each of the extracted factors fits the data from the experimental results. From the analysis results, the main factor affecting the sorting techniques was the time taken to sort. It contributed 98.97%, 98.75% and 94.68% for Heap, Treap and Shell respectively. The Memory consumed came second contributing 1.03% for Heap, 1.25% for Treap and 5.32% for Shell. Total memory consumed was the least of the factors contributing negligible percentages for the three sorting techniques.

In summary, 'time taken' to sort is the main factor affecting the efficiency of the sorting techniques. It was observed that for small data set, shell sort performs better than both heap and treap sort. For small dataset, treap sort has the worst performance, but as the dataset increases, shell running time increases as well more than heap and treap. Since in most real life applications today, dataset are always very large, shell sort does not seem promising. Treap

171

sort averagely has the best performance in terms of running time especially when the data set becomes larger. Also, for memory usage, the treap algorithm used the least memory for operation compared to other algorithms. However, as opposed to the case of running time, shell sort consume less memory when compared to that of heap sort technique. Therefore, it can be concluded that treap sort is a more efficient sorting technique in terms of both running time and memory usage than shell and heap sort most especially when the dataset is very large. Finally , since 'time taken' is to be considered the most paramount factor to users, heap sort performs better than shall so it  will be a better option after treap, otherwise, shell is more efficient. In future work, system environment and software factors could be explored as other factors affecting sorting by these methods. Also, the number of sorting methods could be increased from three to four, five, six or more to get better results.